# EFFECTS OF HUMAN FACTOR ON THE SUCCESS OF INFORMATION TECHNOLOGY OUTSOURCING

Zeinab faraji[1], Neda Abdolvand[2]

1-Department of Social Science and Economics, Alzahra University, Tehran, Iran
2- Assistant Professor, Department of Social Science and Economics, Alzahra University, Tehran, Iran

## ABSTRACT

*Information technology outsourcing is one of the factors affecting the improvement of flexibility and dynamics of enterprises in the competitive environment. Also, the study of the factors affecting its success has been always considered by business owners and the area of research. Professional experiences and research results consider that the success of IT (Information technology) outsourcing projects relates to the effective knowledge transfer and human factors. The human factors are influenced by the cultural and environmental context of the inside and outside of the organization. Hence, it is necessary to study the effectiveness of these variables in different cultural environments. This study investigates the effect of human factors including the customer motivation and vendor willingness on the success of IT outsourcing projects. For this purpose, the research hypotheses were developed and analyzed by the structural equation method. The result of a field study among 94 companies and organizations show the difference of the findings of this study with earlier findings in other countries. Based on the findings, the client motivation doesn't affect the knowledge transfer but the vendor willingness affects the customer motivation to knowledge transfer. This result can help the business owners to take appropriate approaches for achieving success in IT outsourcing projects.*

## 1. INTRODUCTION

Recent advances in technology and increased competition made the companies to attempt for survival and achieving a sustainable competitive advantage. Further, the growth of knowledge has brought toward specialization and increased its managing costs. Outsourcing has been proposed as an approach which can be accompanied with specialized knowledge and reduced costs. Today, outsourcing has increasingly developed and become a service which provides flexibility and dynamism of institutions in a competitive environment. According to Consulting firm Gartner the border of outsourcing is estimated to grow in to a $230 billion by 2015 from $123.6 billion in 2001 [3].

The information technology development and its specialization have led to a dramatic expansion of IT outsourcing to obtain cost savings and operating efficiencies [35]. Outsourcing has long-term and short-term effects on performance and if its long-term benefits be ignored, it may leave many negative effects. Hence, business researchers and practitioners are interested in studying the factors affecting outsourcing success [21][26].

Earlier studies have shown that the effective transfer of technical knowledge can have a significant impact on the effective IT outsourcing. [26], moreover the effective transfer and share of knowledge depends on several factors including: initial trust [22], explicit and tacit knowledge [5], and human factors such as client and service provider capabilities, characteristics, communication and participation that consequently affect the collaborative learning [26][14]. Further, the effectiveness of human factors is affected by culture and environment. Hence, many studies recommended to study the role of human factor in different cultures [3][35] [26][22]. Therefore, the objective of this study is to evaluate the role of human factors on IT outsourcing success in Iranian organizations.

The paper is organized as follow: the study begins with a brief review of the literature and key studies on the management of the role of human resource on the success of knowledge management and IT outsourcing; further the background, and definition of some effective factors on outsourcing is reviewed. Then the research model and hypotheses are explained. The study is followed by a brief review of research methodology, sample and data collection. The last part of the paper includes conclusion and implications of the study.

## 2. LITERATURE REVIEW

### 2.1. Background of Outsourcing

The importance of outsourcing in the business world has been increased in research, which in hence causes to the study of the effective transfer of knowledge [35][21][26][5], effective coordination between client and vendor, and also addressing the role of human relations [22][14] [31] as some of these studies are listed in Table 1.

Table 1: key experimental studies on knowledge management and IT outsourcing

| Resource | Target | Population | Major Findings |
|---|---|---|---|
| [35] | To study the effect of collaboration experiences on knowledge transfer and the performance of IT outsourcing | 146 IT outsourcing partnerships in Singapore. | Human and organizational capabilities accelerate the process of knowledge transfer which leads to more confidence in the IT projects. |
| [26] | To study The role of IT human capability in the knowledge transfer process in IT outsourcing context | vendor and client matched-pair samples of 87 IT outsourcing projects | |
| [21] | To study the effects of knowledge sharing, organizational capability and partnership quality on IS outsourcing success. | IS experts of 195 public sector organizations in Korea | |
| [5] | To study the effect of different kinds of knowledge transfer and sharing on IT outsourcing | IT professionals in banks and companies having outsourcing | |

International Journal of Information Technology Convergence and Services (IJITCS) Vol.6, No.1,February 2016| | | | |
|---|---|---|---|
| | performance. | | |
| [14] | To study the direct and complementary effects of vendor and client IT capabilities on success of IT outsourcing projects | using 267 client–vendor matched data | Human factors are significant factors in flexible relationships and significantly influence outsourcing success. |
| [31] | To Provide integrated framework for the concept of outsourcing success | Professionals and university experts | Lake of specific definition for proposed vocabularies is the main factor that cause to dissatisfaction in outsourcing success. |
| [22] | To measure the effect of trust from the perspective of service provider and receiver on IT outsourcing success | IT professionals of 148 banks | Initial trust affects ongoing affair and has effect on knowledge sharing and outstanding success |

The appearance of IT outsourcing dates back to the early 1960s, when the EDS and Bluecross signed an agreement to manage the services of data processing in EDS [23]. In mid-1980, business of outsourcing has become a pervasive phenomenon and subject of notable research [31]. Moreover, IT has been assigned as an extensive and professional service which can be outsourced. As the extent of IT outsourcing increase, along with its profitability, the number of providers as well as specialized services will increase [23].

Yang and Huang defined outsourcing as delegating the planning, management, implementation and operation of specific tasks to an independent third party [37]. Given the range of IT outsourcing that Yung and Huang suggest, IT outsourcing is defined as significant participation of foreign factor in providing the physical and human resources related to IT in customer organization, or the transfer of assets to various human forms, network or hardware of the customer organization to an external factor which is responsible for delegated activities. Outsourcing happens when a company contracts major goods and services to other companies. Work that is already done internally is shifted to an external service provider, and the employees of the original organization are often transferred to the service provider. There is no profit sharing or mutual contribution in Outsourcing, and it is different with alliance, partnership or joint venture in which resources are transmitted between service provider and client [4]. Park et al. recognized the outsourcing as a well-known and powerful way to shape and organize the management [26]. Cheon et al argue that organizations outsource IT to achieve their objectives and through it obtain information needs from the foreign contractors [6]. Lee stated that the organization, in order to gain strategic, economic and technological advantages may outsource all or part of units' tasks to external service provider [21].

In Yang and Huang's and Bluekert's definition of outsourcing, IT outsourcing relies only on assigning operations [38][4], and Cheon et al focused on the provision of conditions to access information needs [6]. However, Lee focused on the purpose of gaining economic and technological advantages of outsourcing [21].

Regarding these definitions two areas of outsourcing should be considered. The first area is the extent and scope of outsourcing which determines what volume of activities are assigned to an external factor, without jeopardizing the company's survival and intellectual capital. Second area is factors affecting outsourcing success and its objective achievement.

Organizations are considering outsourcing as a factor for gaining economic advantages, increasing flexibility, and improving service quality to better access to new technologies [2]. IT outsourcing influences not only the cost structure of the company, but also the company's long term competition, which changes the nature of risks that the company encounters. Hence, outsourcing brings risks which lack of risk management cause to the company's outsourcing failure [11]. According to earlier research, some risks cause to outsourcing failure including: inaccurate costs estimation, legal challenges, lake of intellectual capital and privacy of information and knowledge, cultural incompatibility, and employee's morale weakness. Unexpected expenses or failing to account the outsourcing total cost especially intangible costs cause to higher costs of outsourcing than expected; further executive disagreement between client and supplier, leads to legal conflicts, which would bring financial and non-financial costs such as negative impact on the reputation and credibility of the company, even to the managers replacement and stock value reduction [2].When an organization outsources its tasks, actually loses its organizational knowledge and it is possible that supplier share or sell the shared information of organization to its competitors [24][33]. Moreover, Cultural differences and biases may also hinder the adoption of provided services by employees and clients or the staff morale may weaken due to outsourcing and lead to lake of obligation and loyalty which can decline performance [24][25]. Other risks are incorrect estimation of projects scheduling and implementation, distance and communication needs, politics and human factors. Error in time estimation is another factor that stem from project managers carelessness which may lead to additional costs and project failure [1]. Despite advanced communication technologies, the time intervals may cause to problems in interaction and transmission of concepts [10]. Political factors are like wind, if is in compliance with the project carry it out and tries to change its direction otherwise. Top management supports, weakness of project management, lack of stakeholder's participation, lack of members' skills are human factors risks. Therefore, study the factors affecting IT outsourcing is essential for risk reduction and success.

Different researches have studied the Outsourcing success factors. Willcocks et al. have categorized the factors into three groups: The first group is related to outsourcing decisions in which the level of outsourcing is referred to the percentage of IT budget or type and number of outsourced IT tasks. These factors along with financial management commitment, the evaluating process of the best suppliers' selection, and human resources related issues are influential. The second group refers to the factors related to the sovereignty of the contract. Which includes four aspects of contract details, type, periods and time of contract, which shows that short term contracts (less than three years) are more successful than long term contracts (more than three years). Third group is related to relationships governance that covers more flexible issues about client and provider relationship management, including trust, norms, open communication, and so on [36]. Since the outsourcing is a standard service without special framework, a deep relationship between provider and client and multiple meetings is necessary [20]. Lee mentioned that knowledge sharing, organizational capability, and partnership quality are effective on outsourcing success. He argued that capturing and sharing of knowledge including both explicit and tacit knowledge increases organization capability. Tacit knowledge is a specific knowledge with unique content, which is difficult for structuring and diffusing, while explicit knowledge is easily distributed in formal and symbolic language [21]. Khan and Khan mentioned that contract flexibility, trustworthy relationship management, competitive bidding, consultation and negotiation, quality management, knowledge sharing, top management support, software process improvement certification, risk sharing attitude, time management, culture awareness, intellectual property right, data security and privacy, detailed specifications of product and project, conflict reconciliation mechanism are the important factors for successful outsourcing partnership, among which first five factors were identified as critical success factors [19].

Also Gonzales et al. recommended a set of factors such as clear vision of outsourcing goal, financial justification, and contract with suitable structure, top managers' support and participation, expert human resources, Assessment of the organization's successful projects as the key factors of outsourcing success. They considered cost saving and suitable infrastructure as two key factors of critical success; however recent studies have been concentrated more on human factors [13]. Theo and Betcherji have introduced human factors such as motivation and ability to communicate and Knowledge codifiability as the knowledge transfer success factors [35]. Han et al. know the human ability as a factor of obtaining IT capabilities, which leads to outsourcing success [14]. Lee and Choi introduced trust and knowledge transfer as human factors of outsourcing success. In many cases, human factor has been studied as having direct role in relationship or indirect role in other factors. With regard to the effect of culture on human factor a most intensive attention to cultures and societies is necessitated [22].

## 2.2 The role of human factors and relationships on outsourcing success

Katz and Kahn defined communication as the information exchange and transfer of meaning. The mutual communication between the service receiver and provider is considered as one of the most effective factors of outsourcing based on cooperation, which creates a win-win relationship [16]. The factor that helps a win-win relationship is regulating an appropriate contract in which all points have been well mentioned. However, there are other factors including trust, commitment, interdependency and knowledge transfer, which despite the significant impact, couldn't properly include in the contracts [22]. Service providers and receivers skill and capability, lead to the accurate relationships and trust between parties; if the providers don't have the appropriate competency and capability to implementation, they lead the project to the failure that causes distrust for the future projects [17]. Moreover, a strong trust causes sharing more knowledge and learning from each other, which leads to knowledge transfer [26].In fact, trust is a key factor in knowledge transfer [27]. Bandyopadhyay and Pathak knew the creation and Share of knowledge as a social process. This Sociability increases the importance of knowledge acquisition by managers. Capability to achieve skills and knowledge transfer which leads to the project's success depends more on individual's motivation and willingness to exchange experiences [3]. Further, human attitudes and factors that are rooted in its character such as honesty, willingness and benevolence has a great impact on trust between parties and knowledge transfer, which can contribute to the project's success. Therefore, role of human factor in the success of outsourcing projects has become more important [35].

With regard to the prominent role of human factor in IT outsourcing and due to the effect of internal, environmental, economical and cultural factors of industry on this role, human factors take more attention in literature. IT outsourcing is important too, and IT outsourcing relationships management is counted as a strategic necessity and enables the organizations to focus on their core tasks to gain more success. Human being has an important role in information exchange and transfer of meaning, and is in upper level than computer and other technologies. One with a true relationship and transfer of knowledge cause to the project success or leads to failure otherwise. Projects related human factor risks are Lack of knowledge, skills and expertise of team members. Some human factors which are influential in knowledge and meaning transfer that result in outsourcing success are shown in table 2 [37].

As mentioned above, the role of human factors and relationships in knowledge transfer is one of the proposed risks of success or failure of projects. Since effective knowledge transfer and share is affected by human factors, the better people interact, the more motivation and incentive they have to cooperate, understand and proper and perfect knowledge transfer. On the other hand, the more this incentive and willingness is, the more people are committed to work, and will show more flexibility to problems. Therefore, individual's incentive and willingness are two important factors regarding the establishment of proper relationship between them which consequently lead to outsourcing success [35][21][26][22] [5].

Table2: Table of examining variables based on literature subject

| Resource | Knowledge transfer | | | | trust | | | Cooperative learning | | |
|---|---|---|---|---|---|---|---|---|---|---|
| | Motivation | willingness | Human ability and competency | Previous experience | Interaction | Supplementary | Human character | Extent of trust | Reciprocal dependency | Promotion of interaction | Procedure of group |
| [21] | * | * | * | * | | | | | | | |
| [22] | | | | | | | * | | | | |
| [26] | * | * | * | | * | | | | * | * | * |
| [14] | | | * | | | * | | | | | |
| [3] | | | * | | * | | | | | | * |

Due to the effect of environment and culture on human factors, this study examines the effect of client motivation and vendor willingness on knowledge transfer and then outsourcing success of Iranian companies is examined.

## 3. RESEARCH MODEL AND HYPOTHESES

Previous studies often examined the effects of human factor on knowledge transfer or the role of knowledge transfer on outsourcing success. With regard to the strategic importance of IT outsourcing relationships management in companies in order to focus more on tasks and achieving more success, we studied these two subjects in line, and the role of human factor on IT outsourcing success is investigated.

- **Client motivation in knowledge transfer**

True flow of knowledge among participant of project is one of the factors that creates trust and true relationships in a project success. Since knowledge sharing in originations is based on background, structure and different goals, knowledge doesn't transfer easily among organizations with different cultures, and there must be common goals between service provider and receiver in order to share knowledge [5].

On the other hand, people that involved in a project don't believe in knowledge sharing and consider it as redundant and bothersome work [28]. Therefore having motivation and motivating people to share proper knowledge is important. Unmotivated receivers can cause to insecurity and passivity in decisions or slyness and subversion in knowledge acceptance and accomplishment [34][30].

Park et al. believed that IT human skills will increase the mutual trust [26]. Client motivation to knowledge acquisition in an IT outsourcing situation is imperative and could be an essential factor for enhancing knowledge transfer. In fact, motivational factors are keys to identify knowledge transfer [35].

Thus:
Hypothesis1. Client motivation is positively related to the extent of knowledge transfer from the vendor.

- **Vendor willingness to knowledge transfer**

In addition to client motivation, the vendors should be willing to share their knowledge with clients. In Knowledge-based industries, such as consulting and software development, many vendors may be reluctant to share their knowledge with service receivers to maintain their competitive advantage [35]. The providers' capabilities [26] and willingness [35] will affect the knowledge transfer.

Thus:
Hypothesis2. Vendor willingness is positively related to the extent of knowledge transfer from the vendor.

- **Knowledge transfer and outsourcing success**

Knowledge of outsourcing can increase IT flexibility [35]. An organization must be able to use its previous knowledge to take advantages of knowledge sharing. In outsourcing relationships, much emphasize is on cooperative relationships than contractual relationships as well as knowledge sharing through this cooperation [21].

Although interactions between involved parties in IT outsourcing has an important role on the extent of knowledge sharing, the value that correct process of knowledge sharing brings to the business leads to operational efficiency of projects and ultimately cause to the success of outsourced projects [5]. Lee stated that by knowledge sharing between service receiver and provider, they will be able to maintain an effective outsourcing relation during the project [21].

Thus:
Hypothesis3. Knowledge transfer is positively related to IT outsourcing success.

## 4. METHOD

This study aims to determine the relationship between IT outsourcing success and knowledge transfer, and the structural equation of this relationship. This study is a fundamental research, regarding the specific objective of the paper and since it tries to identify the effect of human factors on IT outsourcing success, is considered as a survey by its nature. Descriptive study includes methods that aim to describe under study situations or factors. Descriptive study can only used to recognize current conditions or help decisions [29], and to determine the characteristics of variables in a given situation [32]. One case of using descriptive study is to understand the relationship between variables and solving structural equations and factors that this research is based upon [29].

International Journal of Information Technology Convergence and Services (IJITCS) Vol.6, No.1,February 2016## 4.1. Instrument And Data Collection

The instrument of gathering data was questionnaire, which was adjusted in four parts based on 5-point Likert scale that is shown in Table 4. The reliability and validity of questionnaires is examined using experts' ideas and Cranach's alpha, which was 0.80 for the whole scale of questionnaire. Cronbach suggests reliability of 45% as less, 75% as moderate and acceptable, and 95% as high [8]. The Cronbach's alpha of factors was at an acceptable level (see Table 3).
The statistical population of the study includes 150 companies in Tehran that had at least one IT outsourcing task. The questionnaires were distributed among mentioned organizations and 94 questionnaires were collected; that had an optimal return rate of 62/67%.

With regard to the purpose, the structural equation based on mentioned factors was used and analyzed using a Partial Least Squares (PLS) method and Smart PLS software. This technique provides the chance to study the relations between hidden variable and observable variables in parallel. This method is used when the sample size is small or distribution of variables isn't normal [12][15][37].

## 4.2. Results

This study examined the reliability of the model using factor loading coefficient, composite reliability and AVE. Dong et al introduced 0/5 as the adequate AVE [9]. The results indicate that all factors are acceptable at this point (see Table 3). Moreover, in order to assess the models' fitness, coefficient of determination ($R^2$) was used. According to Chin, $R^2$ over 0/67 is strong, over 0/33 is acceptable and 0/14 is poor [7].

According to Klein the desirable factor loading is more than 0/6, factor loading between 0/3 and 0/6 is appropriate and less than 0/3 is poor [18]. The results showed that the loadings of 3 factors (i.e., 7, 8 & 18) were less than 0/5 and were eliminated and the model was re-run. (See Table 5). Composite reliability index must be greater than 0.7 [9]. Hence, this model is verified in terms of reliability and validity.

Based on the results, 33 % of studied organizations were financial, 24 % were in the field of reduction, 21% architecture, and 18% healthcare. In fact, the greatest frequency of studied organizations is related to financial organizations.

Table3. Descriptive Statistics, Cranach's Alphas & Factor loading

| Factor | Number of items | Resource | Cronbach's alpha | Factor loading | | Ave | Composite reliability | R2 |
|---|---|---|---|---|---|---|---|---|
| **Client motivation** | 4 | [35] | 0.80 | Item1 | 0.75 | 0.60 | 0.85 | |
| | | | | Item2 | 0.77 | | | |
| | | | | Item3 | 0.90 | | | |
| | | | | Item4 | 0.67 | | | |
| **Vendor willingness** | 3 | | 0.90 | Item5 | 0.96 | 0.90 | 0.96 | |
| | | | | Item6 | 0.97 | | | |
| | | | | Item7 | 0.06 | | | |
| **Knowledge Transfer** | 4 | | 0.70 | Item8 | 0.40 | 0.70 | 0.85 | 0.40 |
| | | | | Item9 | 0.78 | | | |

| | | | | Item10 | 0.88 | | | |
|---|---|---|---|---|---|---|---|---|
| | | | | Item11 | 0.77 | | | |
| **IT Outsourcing Success** | 9 | [21] | 0.80 | Item12 | 0.75 | 0.50 | 0.86 | 0.21 |
| | | | | Item13 | 0.6 | | | |
| | | | | Item14 | 0.53 | | | |
| | | | | Item15 | 0.61 | | | |
| | | | | Item16 | 0.75 | | | |
| | | | | Item17 | 0.58 | | | |
| | | | | Item18 | 0.47 | | | |
| | | | | Item19 | 0.69 | | | |
| | | | | Item20 | 0.71 | | | |

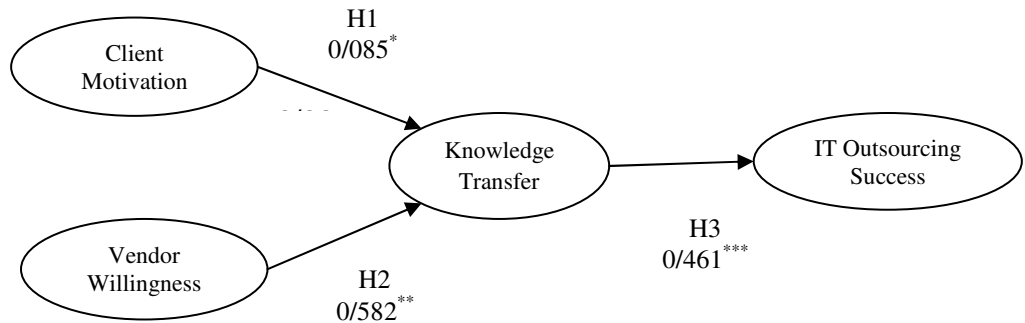

T value=0.803, P value=0.422[*]
T value=6.148, P value=0.000[**]
T value=7.384, P value=0.000[***]

Figure1. Analysis Results

After the model approval, T correlation coefficient and standardized load factor related to pathways of each original structure was used for checking each hypothesis. As it is shown in Table 6, T significant coefficients were greater than 96/1 and standardized factor loading related to each original structure was higher than 0/5, which indicates the importance of the relevant parameters, therefore the hypotheses were confirmed.

## 5. DISCUSSION AND CONCLUSION

Due to significant growth and development of IT and its specialization, IT outsourcing is considered as an effective factor for organizational survival and achieving competitive advantage. Therefore, research that focuses on the study of factors affecting outsourcing success will be of value to research and practice.

Earlier research has been studied the effect of knowledge sharing and transferring index on effective IT outsourcing. Another effective factor is human factor which is influenced by culture and environment. Due to cultural differences, this key factor is being studied in different cultural contexts. Hence, this study examined the effect of human factors including client motivation and vendor willingness on IT outsourcing success in Iranian organizations, which is done by distributing questionnaires among those using these services at least once. According to the results, two hypotheses were confirmed and one of them was rejected. The second and third hypothesizes were confirmed which is in line with previous results such as Theo and Betchurchi that showed the positive effect of vendor willingness on knowledge transfer [35]; and of Bloomberg et al., Lee that indicated the positive effect of knowledge transfer on outsourcing success [21] [5]. While previous studies such as Theo and Betchurchi demonstrated the positive effects of client motivation on the extent of knowledge transfer from the vendor, this hypothesis was rejected in this study [35].

Results suggests that in the studied population, human factor can have an effective role in knowledge transfer; the more vendor is willing to knowledge transfer, the knowledge transfer is in higher level, and seems that the clients motivation has no significant role on knowledge transfer. Therefore, during the contract, we must act in a way that the necessary guarantees for proper and accurate knowledge be transferred by the vendor and its required contexts which could be the clients readiness and time allocation to knowledge acquisition be provided. Besides having administrative knowledge, vendor should be capable of transiting and communicating to knowledge transfer. The other word, it must be ensured that the vendor is willing and capable of transferring knowledge, and it can be consider as an evaluation criteria of the vendor. The more effective knowledge transfer, the greater possibility of success in outsourcing project is.

The major finding of the study is that theories and results of the similar studies in other countries with different cultural contexts cannot be applied in Iran and needs more studies. Moreover, with regard to different cultural context, the theories and results of the research in other countries cannot extend to Iran, and it is better to be tested.

Although, this research studied two factors of client motivation and willingness, focus on these factors is yet the greatest limitation of the study. Therefore future research should study the effects of other factors such as trust, cooperative learning, and interpersonal skills. Moreover future research should study the effective factors on efficient knowledge transfer by vendor. Further, future research should extend the geographical context of the research and do the same study in other industries such as tourism and business.

International Journal of Information Technology Convergence and Services (IJITCS) Vol.6, No.1,February 2016[22] Lee, Jae-Nam, and Byounggu Choi ,(2011), "Effects of initial and ongoing trust in IT outsourcing: A bilateral perspective." Information & Management, Vol 48, No. 2, pp.96-105.
[23] Lee, J.-N., et al, (2003), "IT outsourcing evolution---: past, present, and future." Communications of the ACM, Vol 46, No. 5, pp. 84-89.
[24] Metters, Rich. (2007), "A typology of offshoring and outsourcing in electronically transmitted services", Operations and Management, Vol 26, pp.198-211.
[25] Nakatsu, Robbie T., and Charalambos L. Iacovou. (2009), "A comparative Study of important risk factors involved in offshore and domestic outsourcing of software development projects : A two-panel Delphi study." Information and Management, Vol.46, pp. 57–68.
[26] Park, Joo Yeon, Kun Shin Im, and Joon S. Kim (2011), "The role of IT human capability in the knowledge transfer process in IT outsourcing context." Information & Management, Vol 48, No. 1, pp. 53-61.
[27] Rahnemood ahan, F, Khavndkar, G, (1387), "The effect of knowledge sharing on the successful outsourcing of IT services", IT Management, Vol 1, No.1, pp. 49- 64 (in Persian)
[28] Riege, Andreas (2007), "Actions to overcome knowledge transfer barriers in MNCs." Journal of knowledge management, Vol 11, No. 1, pp.48-67.
[29] Sarmad, Zohreh, Bazargan, Ali. and Hijazi, Emad., (1382), "Research methods in the behavioral sciences", 7th Ed, Tehran: Agah Institute (in Persian)
[30] Schilling, Jan, and Annette Kluge (2009). "Barriers to organizational learning: An integration of theory and research." International Journal of Management Reviews, Vol 11, No. 3, pp.337-360
[31] Schwarz, Colleen (2014), "Toward an understanding of the nature and conceptualization of outsourcing success." Information & Management, Vol.51, No. 1. pp.152-164.
[32] Skaran, U, (2004), "Research Methods in Management", translated by Saebi, M and Shirazi, Mahmoud M (1381), Tehran, Training and Research Planning and Management Institution, 2nd Ed (in Persian)
[33] Sinha, D. and R. Terdiman, (2002), "Potential risks in offshore sourcing." Gartner Group Market Analysis ITSV-WW-DP-0360
[34] Szulanski, Gabriel. (1996), "Exploring internal stickiness: Impediments to the transfer of best practice within the firm." Strategic management journal, Vol 17(S2), pp.27-43.
[35] Teo, Thompson SH, and Anol Bhattacherjee (2014), "Knowledge transfer and utilization in IT outsourcing partnerships: A preliminary model of antecedents and outcomes." Information & Management, Vol 51, No. 2, pp.177-186.
[36] Willcocks, Leslie P., and Thomas Kern. (1998). "IT outsourcing as strategic partnering: the case of the UK Inland Revenue". European Journal of Information Systems, Vol 7, No.1 pp. 29-45.
[37] Wold, Svante. (1993), "Discussion: PLS in chemical practice." Technimetrics, Vol 35, No. 2, pp.136-139.
[38] Yang, Chyan, and Jen-Bor Huang (2000), "A decision model for IS outsourcing". International Journal of Information Management, Vol 20, No.3, p. 225-239.
12